\documentclass[twocolumn]{IEEEtran}

\IEEEoverridecommandlockouts
\usepackage[cmex10]{amsmath}
\usepackage{amssymb}
\usepackage{graphicx}
\usepackage{cite}
\usepackage{enumerate}
\usepackage{epsfig}

\newtheorem{theorem}{Theorem}
\newtheorem{lemma}{Lemma}
\newtheorem{remark}{Remark}
\newtheorem{col}{Corollary}

\begin{document}
\title{Capacity of the State-Dependent Wiretap Channel: Secure Writing on Dirty Paper}
\author{Hamid G. Bafghi, Babak Seyfe, Mahtab Mirmohseni, Mohammad Reza Aref
\thanks{Hamid G. Bafghi and Babak Seyfe are with the Information Theoretic Learning System Lab. (ITLSL), Department of Electrical Engineering, Shahed University, Tehran, Iran. All the authors are with the Information Systems and Security Lab. (ISSL), Deptartment of Electrical Engineering, Sharif University of Technology, Tehran, Iran. Emails: \{ghanizade, seyfe\}@shahed.ac.ir, \{mirmohseni, aref\}@sharif.edu. This work was partially supported by Iran National Science Foundation (INSF), under contracts' numbers 91/s/26278 and 92/32575.}}

\maketitle\pagenumbering{arabic}\pagestyle{empty}\pagestyle{plain}
\begin{abstract}
In this paper we consider the State-Dependent
Wiretap Channel (SD-WC). As the main idea, we model the
SD-WC as a Cognitive Interference Channel (CIC), in which the primary receiver
acts as an eavesdropper for the cognitive transmitter's message.
By this point of view, the Channel State Information (CSI) in SD-WC plays the role of the
primary user's message in CIC which can be decoded at the eavesdropper.
This idea enables us to use the main achievability approaches of CIC,
i.~e., Gel'fand-Pinsker Coding (GPC) and
Superposition Coding (SPC), to find new achievable equivocation-rates
for the SD-WC. We show that these approaches meet the
capacity under some constraints on the rate of the channel
state. Similar to the dirty paper channel, extending the
results to the Gaussian case shows that the GPC
lead to the capacity of the Gaussian
SD-WC which is equal to the capacity of the wiretap channel
without channel state.
Hence, we achieve the capacity of the Gaussian SD-WC
using the dirty paper technique. Moreover, our
proposed approaches provide the capacity of the Binary SD-WC.
It is shown that the capacity of the Binary SD-WC is equal to
the capacity of the Binary wiretap channel without channel
state.
\end{abstract}
\section{Introduction}
Secure communication from an information theoretic perspective
attracts some attentions nowadays~\cite{bibi23, bibi21, bibi69, bibi79, bibi76}.
There are a lot of works to study the secrecy problem in different channel
models, which are inspired of the wiretap channel as a basic
physical layer model~\cite{bibi7}. All these attempts are based on
two principal elements: the \textit{Equivocation} as a measurement
to evaluate the secrecy level at the eavesdropper which is
introduced by Shannon~\cite{bibi28}, and the \textit{Random Coding}~\cite{bibi7} as
a coding scheme which leads to the secrecy condition
in the wiretap channels (see~\cite{bibi76} and the references therein).

Using the Channel State Information (CSI) in information theoretic communication
models was initiated by Shannon~\cite{bibi39} in which he
considered the availability of CSI at the Transmitter (CSIT).
Gel'fand and Pinsker obtained the capacity of the discrete
memoryless channel with non-causal CSIT~\cite{bibi22}. Their main
result was based on a binning scheme named Gel'fand-
Pinsker Coding (GPC), and it was shown that the capacity of
the Gaussian state-dependent channel is
equal to the capacity of a channel with no channel state~\cite{bibi30}.
This means that the transmitter, who
knows the CSI non-causally, can adapt its signal to the channel
state such that the receiver senses no interference~\cite{bibi30}, as if the
receiver had knowledge of the interference and could subtract
it out.
The name of the \textit{Dirty Paper} channel~\cite{bibi30} is
inspired of a spotted paper, one wish to write on such that
the reader can easily find the text.

Despite of all these works
which are trying to cancel out the channel state, the authors
in~\cite{bibi75} and~\cite{bibi77} deal with the CSIT in an innovative manner. In these
works, the CSIT assumed as the user's signal which is wished
to be sent through the channel.
In~\cite{bibi77}, the transmitter wishes
to mask the CSI at the legitimate receiver, whereas in~\cite{bibi75} the
transmitter wishes to forward the CSI to the legitimate receiver.
The achievable rate in each case is derived and
results to the trade-off between the rates of the transmitter's information and
the CSI.

The State-Dependent Wiretap Channel (SD-WC) was studied
in~\cite{bibi23, bibi21}. In~\cite{bibi23}, the authors derived an achievable
equivocation-rate for the SD-WC in general case, and in~\cite{bibi21}
the Gaussian SD-WC was considered. The main ideas
of these papers were based on using a combination of GPC and random
coding. In these works, it was assumed that the eavesdropper
has no access to the CSI. Therefore, the CSI can be used potentially to improve
the secrecy rate of the SD-WC. Specially in Gaussian case, under some constraints,
the capacity was derived as the capacity of a channel~\cite{bibi23, bibi21} with no state.
Afterward,~\cite{bibi40} studied the SD-WC from the
secret-key sharing aspect. In this model, the channel state was
assumed as a key to achieve the secrecy rate. El-Gamal et.~al.,
improved the equivocation-rate for the SD-WC in the case
that the channel state is known causally at both sides~\cite{bibi42}.

\begin{figure}
\centering
\epsfig{file=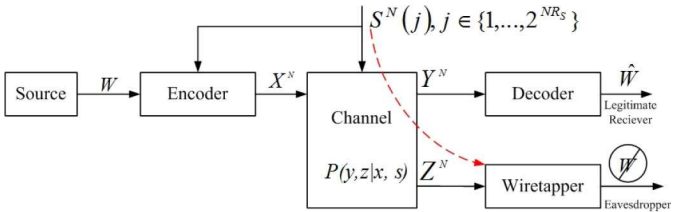,width=1\linewidth,clip=}
\caption{The State-Dependent Wiretap Channel (SD-WC) in which the channel
state is assumed to be known non-causally at the transmitter.}
\label{fig:1}
\end{figure}

In this paper we study the SD-WC (see Fig.~\ref{fig:1}). We consider
the channel state sequence as a random sequence with elements
drawn from a finite alphabet, known non-causally at the
transmitter. The transmitter wishes to keep its message secret
from an external eavesdropper. A well known concept is
to cancel out the CSI using the GPC and use the random
coding to achieve the secrecy rate as~\cite{bibi23}. We know from
the random coding idea~\cite{bibi7} that we should randomize a part
of the message to confuse the eavesdropper which has less
channel capacity than the main channel. In this paper,
we model the CSI of SD-WC as the message of the primary user at CIC and we use it at
the transmitter to randomize a part of the message.
Our main  idea relies on modeling the CSI with a (primary) message, since both are known at the (cognitive) transmitter~\cite{bibi31, bibi9}.
In this model, the transmitter of SD-WC plays the role of the
cognitive transmitter of CIC and the CSI is considered as
the message of the primary one.
The transmitter, similar to the cognitive one in CIC, tries to cancel out the CSI,
equivalently the primary message, at its destination with low enough information
leakage of its message to the eavesdropper, i.~e., the primary receiver in CIC.
Then, we use the rate splitting and the interference cancelation schemes as
CICs in~\cite{bibi31, bibi9}, i.~e., GPC and Superposition Coding (SPC),
to achieve the secrecy rate for the SD-WC. At first, using
these two coding schemes, we derive two new achievable
equivocation-rates for the SD-WC. Then, we prove that these
achievable equivocation-rates meet the secrecy capacity under
some constraints. As an example, we show that the derived
equivocation-rates meet the capacity in the binary SD-WC.
Afterward, we use these results in the Gaussian SD-WC. We prove that
the achievable equivocation-rates for the SD-WC lead to the
secrecy capacity using the \textit{Dirty Paper Coding} (DPC)~\cite{bibi30}
approach.
On account of this result, we call this approach as
\textit{secure writing on dirty paper} to point to the
problem of writing a message on a sheet of paper covered with
independent dirt spots, such that the reader can distinguish the
message, but the eavesdropper cannot do that.

The rest of the paper is organized as follows. In Section II
the channel model is introduced. In Section III the encoding
strategy for the point to point state-dependent communication
channel is explained.
The main results on the achievable equivocation-rates,
the capacity of the SD-WC and the binary example
are presented in Section IV.
In Section V, using DPC, the capacity of the Gaussian SD-WC is derived
and the paper is concluded with some discussions in the
last section.

\section{Channel Model, Definitions and Preliminaries}\label{S2}
First, we clear our notation in this paper. Let~$\mathcal{X}$ be a
finite set. Denote its cardinality by~$|\mathcal{X}|$. If we consider~$\mathcal{X}^N$,
its members are~$x^N = (x_1, x_2,\ldots, x_N)$, where subscripted
letters denote the components and superscripted letters denote
the vector. A similar convention applies to random vectors
and random variables, which are denoted by uppercase letters.
Denote the zero mean Gaussian random variable with
variance~$P$ by~$\mathcal{N}(0, P)$. A Bernoulli random variable~$X$
with~$Pr\{X = 1\} = P$ is denoted by~$X \sim \mathcal{B}(P)$.

Consider the channel model shown in Fig.~\ref{fig:1}. Assume that
the state information of the channel, i.e.,~$S_i,~ 1\leq i \leq N$ is
known at the transmitter non-causally. The transmitter sends
the message~$W$ which is uniformly distributed on the set~$\mathcal{W} \in
\{1,\ldots, M\}$ to the legitimate receiver in~$N$ channel uses. Based
on the~$w$ and~$s^N$, the transmitter generates the codeword~$x^N$
to transmit through the channel. The decoder at the legitimate
receiver makes an estimation of the transmitted message~$\hat{w}$
based on the channel output~$y^N$; and~$z^N$ is the corresponding
output at the eavesdropper. The channel is memoryless, i.e.,
\begin{IEEEeqnarray}{rCl}\label{eqn1}
p (y^{N}, z^{N}|x^{N}, s^{N})
=\prod_{i=1}^{N} p(y_{i}, z_{i}|x_{i}, s_{i}).
\end{IEEEeqnarray}

The secrecy level of the transmitter's message at the eavesdropper is measured by normalized equivocation:
\begin{IEEEeqnarray}{rCl}\label{eqn2}
R_e^{(N)}=\frac{1}{N} H(W|Z^N, S^N),
\end{IEEEeqnarray}
in which we assume that the eavesdropper can decode the CSI.
The average probability of error~$P_e$ is given by
\begin{IEEEeqnarray}{rCl}\label{eqn3}
P_e=\frac{1}{M}\sum_{i=1}^{M}Pr(\hat{w}(Y^N)\neq w).
\end{IEEEeqnarray}

We define the rate of the transmission to the intended receiver to be
\begin{IEEEeqnarray}{rCl}\label{eqn4}
R=\frac{\log M}{N}.
\end{IEEEeqnarray}
The rate-pair~$(R,R_e)$ is achievable if for any~$0\leq\epsilon\leq 1$ there exists
an~$(M,N, P_e)$ code such that~$P_e\leq \epsilon$, and the secrecy rate~$R_e$ is
\begin{IEEEeqnarray}{rCl}\label{eqn5}
0 \leq R_{e}\leq \liminf_{N\rightarrow \infty} R_{e}^{(N)}.
\end{IEEEeqnarray}

We define the \textit{perfect secrecy condition} as the case in which
we have~$R \leq R_e$. Thus, under this condition, the sent message to the legitimate receiver must be secure.

\section{Overview of Encoding Strategy}\label{S3}
Consider SD-WC in Fig.~\ref{fig:1}. Assume that the channel
state~$S$ plays the role of the codebook of rate~$R_S =
I(S; Y )$ interfering with the communication of message~$ W$
at the rate~$ R = I(X; Y )$. As proposed in~\cite{bibi13}, two coding strategy
can be used for this scenario: GPC scheme and SPC depending on the interference's
rate~$R_S$. When~$R_S$ is small, it can be exploited for achieving
the higher rates using the SPC. Using the GPC, the interference
is considered as a codebook with rate~$R_S$.
The following lemma expresses the result of using these two
coding schemes. This lemma is used to derive the achievable
equivocation-rate for the SD-WC in the next section.

\begin{lemma}{~\cite[Lemma 1]{bibi13}}
The following rate is achievable for a point to point communicating system with non-causal CSIT
\begin{IEEEeqnarray}{rCl}\label{eqn6}
R &\leq& \max_{P_{U|S}, x(u,s)}\min\{ I(X; Y |S),\IEEEnonumber\\
&&\max\{I(U, S; Y ) - R_S, I(U; Y ) - I(U; S)\}\}.
\end{IEEEeqnarray}
\end{lemma}
\begin{IEEEproof}[Outline Of the Proof]
This lemma is proved in [17, Appendix
B]. For the case~$I(S;U,Y) \leq R_S \leq H(S)$, the second
term of~\eqref{eqn6} is achievable using the GPC. For~$R_S \leq I(S;U, Y )$,
the first term in~\eqref{eqn6} is achievable using the SPC.
\end{IEEEproof}

\begin{remark}
Interestingly, in the case~$R_S \leq I(S;U, Y )$, the
receiver decodes both the messages and the channel state, i. e.,
the interference. A related model in which both data and
the channel state are decoded at the receiver was
considered by~\cite{bibi75}.
\end{remark}
\section{Results on The Capacity of The SD-WC}
In this section, we provide two achievable equivocation-rates for the SD-WC,
shown in Fig.~\ref{fig:1}, in two cases, i.~e., GPC and SPC.
\begin{figure}
\centering
\epsfig{file=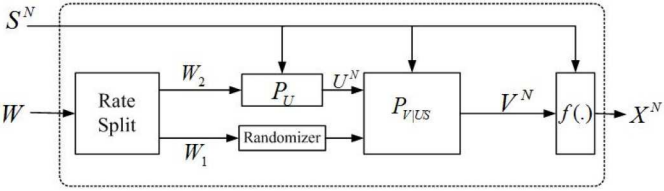,width=1\linewidth,clip=.5}
\caption{Gel'fand-Pinsker Coding (GPC) scheme for the
case~$\min\{I(S;U; Y ); I(S; V; Y |U)\} \leq R_S \leq H(S)$.}
\label{fig:2}
\end{figure}

\subsection{Gel'fand-Pinsker Coding (GPC)}
Assume that~$V$ and~$U$ are two random variables to construct
the private and the common messages, respectively. In the case
that~$\min\{I(S;U, Y ), I(S; V, Y |U)\} \leq R_S \leq H(S)$, using
GPC and the rate splitting, we have the following lemma for
the achievable equivocation-rate of the SD-WC.

\begin{lemma}[Achievable equivocation-rate using GPC]\label{lem2}
The set of rates~$(R,R_1,R_2,R_e)$ satisfying
\begin{IEEEeqnarray}{rCl}\label{eqn7}
\mathcal{R}^{GPC}&=&\bigcup_{P(T)P(S,U,V,X|T)P(Y,Z|X,S)}\IEEEnonumber\\
&&\left\{\begin{array}{l}
(R,R_1,R_2,R_e) :\\
R = R_1 + R_2\\
R_1\leq I(V ; Y |U, T)  - I(V ; S|U, T),\\
R \leq  I(V,U; Y |T)    - I(V,U; S|T),\\
R_e\leq  I(V ; Y | U, T) - I(V ; S,Z | U, T),\\
\end{array}\right\},\IEEEnonumber\\
\end{IEEEeqnarray}
is achievable for SD-WC, in which the equations in the right-hand-sides
of~\eqref{eqn7} are non-negative and~$T$ is a time-sharing random
variable.
\end{lemma}
\begin{IEEEproof}[Outline of the Proof]
The details of the proof are relegated to the Appendix A based on the result
of~\cite{bibi31}. As an outline, the message~$W$ is split into
two messages~$W_1$ and~$W_2$ with denoted rates~$R_1$ and~$R_2$,
respectively. In our scheme,~$W_2$ can be decoded at the eavesdropper
and does not contribute to the secrecy level of~$W$ at the
eavesdropper. Therefore,~$W_1$ may be hidden from the unintended
receiver. We model SD-WC with
a CIC with a confidential message~\cite{bibi31}, in which the message of the primary transmitter
plays the role of CSI and the transmitter who
knows CSI, can be considered as the cognitive transmitter.
By this setting, the proof of Lemma~\ref{lem2} is deduced from the
proof of~\cite[Theorem 1]{bibi31} in which a cognitive radio tries
to communicate with a related destination through a main
channel which belongs to a primary transmitter-receiver pair.
Therefore, the transmitter in SD-WC (similar to the cognitive transmitter in CIC),
can cancel the channel state out at the corresponding
receiver, using the known CSI. The transmitter uses rate splitting scheme
and GPC by binning~$U$ against the channel state~$S$, and
binning~$V$ against~$S$ conditioned on~$U$. The GPC structure
is shown in Fig.~\ref{fig:2}.

\end{IEEEproof}

The perfect secrecy condition using the GPC scheme is
derived as follows.
\begin{theorem}[Perfect secrecy condition using GPC]\label{thm1}
In the case that $\min\{I(S;U, Y ), I(S; V, Y |U)\} \leq R_S \leq H(S)$, the
perfect secrecy rate~$R^{GPC}_{ps}$ satisfying
\begin{IEEEeqnarray}{rCl}\label{eqn8}
\mathcal{R}^{GPC}_{ps}&=&\bigcup_{P_{U,V,S}P_{X|U,V,S}P_{Y,Z|X,S}}\IEEEnonumber\\
\Big\{R_e &\leq& \min\{I(V,U; Y ) - I(V,U; S),\IEEEnonumber\\
&&I(V ; Y |U) - I(V ; S,Z|U)\}\Big\},
\end{IEEEeqnarray}
is achievable for the SD-WC.
\end{theorem}

\begin{IEEEproof}
The proof of the theorem is strictly deduced from
Lemma~2 applying perfect secrecy condition and using Fourier-Motzkin
elimination~\cite{bibi69}.
\end{IEEEproof}

\begin{col}
The following rate
\begin{IEEEeqnarray}{rCl}\label{eqn9}
\mathcal{R}&=& \max_{P_{U,V,S}P_{X|U,V,S}P_{Y,Z|X,S}}
I(V ; Y |S) \IEEEnonumber\\
&&- \max\{I(V ;Z|S),H(S|Y )\},
\end{IEEEeqnarray}
is an achievable secrecy rate for the SD-WC.
\end{col}
\begin{IEEEproof}
Substituting~$U = S$ in the proof of the Corollary~1 means that the transmitter sends the channel state~$S$ as
a part of its message. Thus, the channel state can be decoded
at the legitimate receiver. This setting is similar to the one
proposed by~\cite{bibi75}, in which the transmitter sends its data and
the information about the channel state to the receiver, without secrecy issue.
\end{IEEEproof}

\begin{remark}

The proof of the corollary is deduced directly from
Theorem 1 by substituting~$U = S$ and using Fourier-Motzkin
elimination~\cite{bibi69}. Note that the transmitter which knows CSI non-causally
can use the channel state ransom variable~$S$ as part of its message, i.~e., the common one.

\end{remark}
\begin{figure}
\centering
\epsfig{file=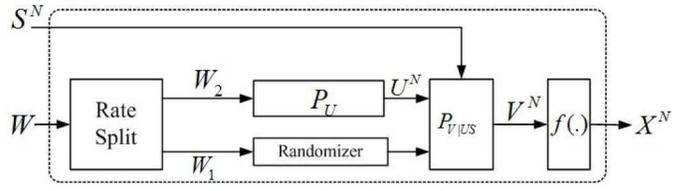,width=1\linewidth,clip=.5}
\caption{Superposition Coding (SPC) scheme for the case~$R_S \leq
\min\{I(S;U; Y ); I(S; V; Y |U)\}$.}
\label{fig:3}
\end{figure}

\subsection{Superposition Coding (SPC)}
Consider the case~$R_S \leq \min\{I(S;U, Y ), I(S; V, Y |U)\}$.
From Lemma~1, we can superimpose~$U^N$ and~$V^N$
against~$S^N$ instead of using GPC. To derive the perfect secrecy rate
for the SD-WC in this case, first we present the following
lemma.
\begin{lemma}[Achievable equivocation-rate using SPC]\label{lem3}
In the case that $R_S \leq \min\{I(S;U, Y ), I(S; V, Y |U)\}$, the
following equivocation-rate is achievable for the SD-WC:
\begin{IEEEeqnarray}{rCl}\label{eqn10}
\mathcal{R}^{SPC}&=&\bigcup_{P_{U,V,S}P_{X|U,V,S}P_{Y,Z|X,S}}\IEEEnonumber\\
&&\left\{\begin{array}{l}
(R, R_e) :\\
R  \leq I(U, V ; Y |S)\\
R_e \leq I(V ; Y |U, S)- I(V ;Z|U, S)\\
\end{array}\right\}.\IEEEnonumber\\
\end{IEEEeqnarray}
\end{lemma}

\begin{IEEEproof}[Outline of the Proof]
The transmitter splits its message~$W$
into two components~$W_1$ and~$W_2$ with denoted
rates~$R_1$ and~$R_2$, respectively. Modeling SD-WC with a
CIC with confidential message~\cite{bibi9}, the proof of the lemma
is deduced from the proof of the~\cite[Theorem 1]{bibi9} in which
the cognitive transmitter superimposes its message on the
primary transmitter's message. In our approach, the transmitter
superimposes its message on the channel state sequence~$S$ to
derive the achievable equivocation-rate for the SD-WC. The
encoding structure in this case is shown in Fig.~\ref{fig:3}. More details
on the proof are relegated to the Appendix B.
\end{IEEEproof}

Thus, using the superposition coding scheme, the following
theorem presents the perfect secrecy condition for the SD-WC.

\begin{theorem}[Perfect secrecy condition using SPC]\label{thm2}
In the case~$R_S \leq \min\{I(S;U,Y), I(S; V, Y |U)\}$, the perfect
secrecy rate~$R^{SPC}_{ps}$ satisfying
\begin{IEEEeqnarray}{rCl}\label{eqn11}
\mathcal{R}^{SPC}_{ps}&=&\bigcup_{P_{U,V,S}P_{X|U,V,S}P_{Y,Z|X,S}}\IEEEnonumber\\
&&\Big\{R_e \leq I(V ; Y |U, S) - I(V ;Z|U, S)\}\Big\},
\end{IEEEeqnarray}
is achievable for SD-WC.
\end{theorem}

\begin{figure}
\centering
\epsfig{file=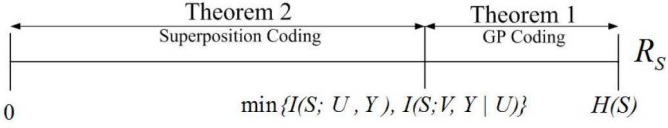,width=1\linewidth,clip=.5}
\caption{The results of the theorems 1 and 2, respect to the channel state
information rate ~$R_S$.}
\label{fig:4}
\end{figure}

\begin{IEEEproof}
The proof is strictly deduced from
the Lemma~\ref{lem3} applying perfect secrecy condition and using Fourier-
Motzkin elimination.
\end{IEEEproof}
Fig.~\ref{fig:4} shows the conditions of using GPC and SPC
in theorems~1 and~2 respect to the CSI rate~$R_S$.

\begin{col}
The rate
\begin{IEEEeqnarray}{rCl}\label{eqn12}
\mathcal{R}=\max_{P_{U,V,S}P_{X|U,V,S}P_{Y,Z|X,S}}
\{I(V ; Y |S) - I(V ;Z|S)\},
\end{IEEEeqnarray}
is an achievable secrecy rate for the SD-WC in the case that~$R_S\leq
\min\{I(S;U, Y ), I(S; V, Y |U)\}$.
\end{col}

\begin{IEEEproof}
The proof of the corollary is deduced directly from
Theorem 2 by substituting~$U =\emptyset$ and using Fourier-Motzkin
elimination~\cite{bibi69}.
\end{IEEEproof}

\subsection{Capacity of the SD-WC}
Now, we have the following theorem to explain the secrecy
capacity of the SD-WC in general case.
\begin{theorem}[Secrecy Capacity]\label{thm3}
In the cases that
\begin{IEEEeqnarray}{rCl}\label{eqn13}
R_S \leq \min\{I(S;U, Y ), I(S; V, Y |U)\},
\end{IEEEeqnarray}
or
\begin{IEEEeqnarray}{rCl}\label{eqn14}
R_S\geq\max\Big\{\min\{I(S;U, Y ), I(S; V, Y |U)\},\IEEEnonumber\\
I(V,U; S) - I(U; Y |S) - I(V ;Z|S)\Big\},
\end{IEEEeqnarray}
the secrecy capacity~$(C_{S})$ of the SD-WC is as follows
\begin{IEEEeqnarray}{rCl}\label{eqn15}
C_{S}&=&\max_{P_SP_{U,V |S}P_{X|U,V,S}P_{Y,Z|X,S}}
\Big\{I(V ; Y |U, S)\IEEEnonumber\\
&&- I(V ;Z|U, S)\Big\}.
\end{IEEEeqnarray}
\end{theorem}

\begin{IEEEproof}
The achievability of~\eqref{eqn15} is derived directly from
Theorem 1 and Theorem 2. In more detail, in the case
that~$R_S \leq\min\{I(S;U, Y ), I(S; V, Y |U)\}$, the rate~\eqref{eqn15} is
achievable for the SD-WC by using SPC as Theorem~\ref{thm2}. In the
case that~$\max\{\min\{I(S;U, Y ), I(S; V, Y |U)\}, I(V,U; S) -
I(U; Y |S)-I(V ;Z|S)\}\leq R_S$, GPC achieves the rate~\eqref{eqn15}, by
using Theorem~\ref{thm1} and substituting~$U = (U, S)$. The converse
proof is relegated to the Appendix C.
\end{IEEEproof}

\begin{remark}
The capacity of the SD-WC in Theorem~\ref{thm3},
in the case that~$U = S = \emptyset $, is reduced to the capacity
of the wiretap channel without channel state information.
\end{remark}

\begin{col}\label{col3}
In the case that~$Y$ is more capable than~$Z$, i.~e.,~$I(X; Y|U)\geq I(X; Z|U)$ for all~$p(x)$,
and under the conditions~\eqref{eqn13}--\eqref{eqn14},
the secrecy capacity of the SD-WC is as follows
\begin{IEEEeqnarray}{rCl}\label{eqn151}
\!\!\!\!\!C_{S}&=&\max_{P_SP_{X|S}P_{Y,Z|X,S}}
\Big\{I(X ; Y | S)- I(X ;Z| S)\Big\}.
\end{IEEEeqnarray}
\end{col}

\begin{proof}
The achievability of~\eqref{eqn151} is directly proved by substituting~$U =\emptyset, V = X $
in Theorem~\ref{thm3}. To prove the converse, we have
\begin{IEEEeqnarray}{rCl}\label{eqn152}
I(V ; Y |U, S)&-& I(V ;Z|U, S)\IEEEnonumber\\
&=& I(X, V ; Y |U, S)- I(X, V ;Z|U, S)\IEEEnonumber\\
&&-[I(X ; Y |U, S, V)- I(X ;Z|U, S, V)]\IEEEnonumber\\
&=& I(X ; Y |U, S)- I(X ;Z|U, S)\IEEEnonumber\\
&&+[I(V;Y|U, S, X)-I(V;Z|U, S, X)]\IEEEnonumber\\
&&-[I(X ; Y |U, S, V)- I(X ;Z|U, S, V)]\IEEEnonumber\\
&\stackrel{(a)}{=}& I(X ; Y |U, S)- I(X ;Z|U, S)\IEEEnonumber\\
&&-[I(X ; Y |U, S, V)- I(X ;Z|U, S, V)]\IEEEnonumber\\
&\stackrel{(b)}{\leq}& I(X ; Y |U, S)- I(X ;Z|U, S),
\end{IEEEeqnarray}
where~$(a)$ is due to the Markov chain relationship~$(U,V)\rightarrow(X,S)\rightarrow(Y,Z)$
which implies that~$[I(V;Y|U, S, X)-I(V;Z|U, S, X)]= 0$; and~$(b)$ is derived by using
the more capable condition. Now, we have
\begin{IEEEeqnarray}{rCl}\label{eqn152}
I(X ; Y |U, S)&-& I(X ;Z|U, S)\IEEEnonumber\\
&=& I(U, X ; Y | S)- I(U, X ;Z|S)\IEEEnonumber\\
&& - [I(U, X ; Y | S)- I(U, X ;Z|S)]\IEEEnonumber\\
&=& I(X ; Y | S)- I(X ;Z|S)\IEEEnonumber\\
&&+ [I(U ; Y | S)- I(U ;Z|S)]\IEEEnonumber\\
&& - [I(U, X ; Y | S)- I(U, X ;Z|S)]\IEEEnonumber\\
&\stackrel{(c)}{\leq}& I(X ; Y | S)- I(X ;Z| S),
\end{IEEEeqnarray}
where~$(c)$ is derived by using the the Markov chain relationship and the more capable condition.

\end{proof}

\subsubsection{An Example (Binary SD-WC)}
As an example, consider
the Binary SD-WC (BSD-WC), in which the channel outputs
are described as
\begin{IEEEeqnarray}{rCl}\label{eqn16}
Y = X \oplus S \oplus \eta_1,\\\label{eqn17}
Z = X \oplus S \oplus \eta_2,
\end{IEEEeqnarray}
where~$\eta_1\sim \mathcal{B}(N_1)$,~$\eta_2\sim \mathcal{B}(N_2)$ and~$S\sim \mathcal{B}(Q)$.
For this channel we have the following theorem.
\begin{theorem}\label{thm4}
When
\begin{IEEEeqnarray}{rCl}\label{eqn18}
R_S \leq I(S; Y ),
\end{IEEEeqnarray}
or
\begin{IEEEeqnarray}{rCl}\label{eqn19}
R_S \geq\max\{I(S; Y ), I(X; S) - I(X;Z|S)\},
\end{IEEEeqnarray}
the secrecy capacity of the BSD-WC is
\begin{IEEEeqnarray}{rCl}\label{eqn20}
C_S^{BSD-WC}=[H(N_2)- H(N_1)]^+,
\end{IEEEeqnarray}
in which~$[x]^+ = \max\{0, x\}$.
\end{theorem}

\begin{IEEEproof}
Substituting~$U =\emptyset, V = X $ in Theorem~\ref{thm3} leads
to the secrecy capacity as following under the conditions~\eqref{eqn18}-\eqref{eqn19}
\begin{IEEEeqnarray}{rCl}\label{eqn21}
C_S^{'}=\max_{P_SP_{X|S}P_{Y,Z|X,S}}
I(X; Y |S) - I(X;Z|S).
\end{IEEEeqnarray}
Let~$X \sim \mathcal{B}(P)$ which is independent of~$S$.
Note that the nature of the BSC forces us to choose Bernoulli distribution function
for the channel input.Now, we have
\begin{IEEEeqnarray}{rCl}\label{eqn22}
&&I(X; Y |S)- I(X;Z|S)\IEEEnonumber\\
&=& H(Y |S) - H(Y |X, S) - H(Z|S) + H(Z|X, S)\IEEEnonumber\\
&=& H(X \oplus \eta_1) - H(\eta_1) - H(X \oplus \eta_2) + H(\eta_2)\IEEEnonumber\\
&=& H(N_2) - H(N_1) - [H(P * N_2) - H(P * N_1)],
\end{IEEEeqnarray}
in which~$P * u = P(1 - u) + (1 - P)u$ and~$H(N_i) =
-N_i \log(N_i) - (1- N_i) \log(1 - N_i)$,~$i = \{1, 2\}$ is the binary entropy function.
For~$P < \frac{1}{2}$ ,
the function~$P *u$ is monotonically increasing in~$u \in [0, 1/2]$.
Hence, setting~$P = 1/2$ we have~$H(P *N_2)-H(P *N_1) = 0$
which achieves the maximum of~$I(X; Y |S) - I(X;Z|S)$
as~$H(N_2)-H(N_1)$. In the case that~$N_1 \leq N_2$ the right hand
side of~\eqref{eqn22} is negative. It means that the eavesdropper can
decode any message intended for the receiver.
The proof of the converse is directly derived from Corollary~\ref{col3}
which implies that the right-hand-side of~\eqref{eqn21} is an outer bound
on the capacity of the model in which the legitimate receiver is more
capable than the eavesdropper. This completes the proof.
\end{IEEEproof}

\begin{remark}
The capacity of the BSD-WC under the conditions~\eqref{eqn18} and~\eqref{eqn19},
is equal to the capacity of the binary wiretap
channel without channel state. It means that the coding
schemes used in the theorems 1 and 2 cancels the channel state out to
meet the capacity.
\end{remark}

\section{Gaussian SD-WC: Dirty Paper Scheme}
\begin{figure}
\centering
\epsfig{file=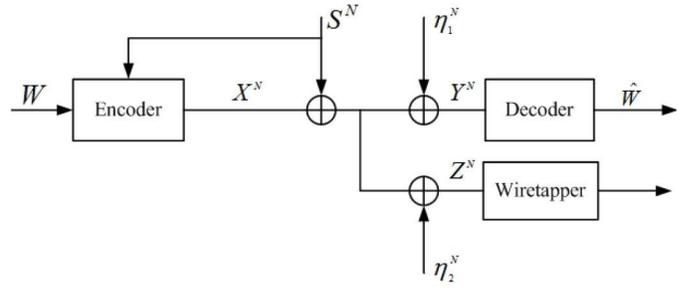,width=1\linewidth,clip=.5}
\caption{Gaussian state-dependent wiretap channel.}
\label{fig:5}
\end{figure}

In this section, we extend the results of the theorems 1
and 2 to the Gaussian SD-WC (GSD-WC). First, consider the
GSD-WC (Fig.~\ref{fig:5}) which is described as follows.

\begin{IEEEeqnarray}{rCl}\label{eqn23}
Y = X + S + \eta_1,\IEEEnonumber\\
Z = X + S + \eta_2,
\end{IEEEeqnarray}
where~$X$ denotes the channel input and~$Y$ and~$Z$ denote the
channel outputs at the legitimate receiver and eavesdropper,
respectively.~$\eta_i \sim N(0,N_i), i \in \{1, 2\}$ is Additive White
Gaussian Noise (AWGN), and we assume the channel state
random variable as~$S \sim N(0,Q)$. Now, we have the following
theorem for the GSD-WC.

\begin{theorem}
The secrecy capacity of the GSD-WC is
\begin{IEEEeqnarray}{rCl}\label{eqn24}
C_S^{GSD-WC}=[\mathcal{C}(\frac{P}{N_1})- \mathcal{C}(\frac{P}{N_2})]^+,
\end{IEEEeqnarray}
in which~$\mathcal{C}(x) = \frac{1}{2} \log(1 + x)$.
\end{theorem}

\begin{IEEEproof}
The proof the achievability of the rate~\eqref{eqn24} is
derived from Theorem 1 and 2 as follows:

First, we assume the channel input as~$X\sim N(0, P)$ which
is independent of the channel state sequence and the AWGNs.
Then, we use the Dirty Paper approach~\cite{bibi30} on theorems~1
and~2, directly. For this, we split the channel input as~$X =
X_1 + X_2 +\sqrt{\bar{\frac{\alpha}P}{Q}}$, in which~$X_1$ is related to the confidential message
and~$X_2$ is related to the common message which reduces the interference at
the legitimate receiver, and we have
\begin{IEEEeqnarray}{rCl}\label{eqn25}
X_1 &\sim& \mathcal{N}(0, \alpha\bar{\beta} P),\\\label{eqn26}
X_2 &\sim& \mathcal{N}(0, \alpha\beta P),
\end{IEEEeqnarray}
where~$0\leq\alpha, \beta \leq 1$ are the power coefficients for sending
the confidential and the common messages in the transmitter,
respectively. Also we define~$\bar{\alpha}=1-\alpha, \bar{\beta}=1-\beta$.

Let the axillary random variables~$V$ and~$U$ as
\begin{IEEEeqnarray}{rCl}\label{eqn27}
V = X_1 + \lambda_1 S,\\\label{eqn28}
U = X_2 + \lambda_2 S
\end{IEEEeqnarray}
in which the transmitter uses~$0\leq\lambda_1, \lambda_2 \leq 1$ to bin its message
against the state of the channel as GPC. Note that in the SPC
the transmitter does not forward the channel state, and the
channel state is not contained in~$V$. Thus, in SPC case, we
should substitute~$\lambda_1 =0$.
Now, we find the variables~$\alpha, \beta, \lambda_1, \lambda_2$ which maximize the
achievable secrecy rate of Theorem 1 leading to the capacity of GSD-WC.

Based on Theorem 1, for the GPC case, we can define
\begin{IEEEeqnarray}{rCl}\label{eqn29}
R_{e_1} &\triangleq& I(V,U; Y ) - I(V,U; S)\IEEEnonumber\\
&=& H(V,U|S) - H(V,U|Y ),\IEEEnonumber\\
R_{e_2} &\triangleq& I(V ; Y |U) - I(V ; S,Z|U)\IEEEnonumber\\
&=& H(V |U) - H(V |Y,U)\IEEEnonumber\\
&&-H(S,Z|U) + H(S,Z|U, V ).
\end{IEEEeqnarray}

Now, we calculate each term of~\eqref{eqn29} using the standard
approach~\cite{bibi34} yielding~\eqref{eqn30}-\eqref{eqn35} in top of the next page.
\begin{figure*}
\begin{IEEEeqnarray}{rCl}\label{eqn30}
H(V,U|S)   &=& \frac{1}{2} \log[(2\pi e)^2\alpha^2 \beta \bar{\beta} P^2],\\\label{eqn31}
H(V,U|Y )  &=& \frac{1}{2} \log\Big[(2\pi e)^2\{(\alpha \beta P+ \lambda_2^2 Q)(\alpha\bar{\beta} P+ \lambda_1^2 Q)- \lambda_1^2 \lambda_2^2 Q^2 +\frac{A}{B}\}\Big],\\\label{eqn32}
H(V |U)    &=& \frac{1}{2} \log\Big[(2\pi e)\{(\alpha\bar{\beta} P+ \lambda_1^2 Q)-\frac{\lambda_1^2 \lambda_2^2 Q^2}{\alpha \beta P+ \lambda_2^2 Q}\}\Big]\\\label{eqn33}
H(V |Y,U)  &=& \frac{1}{2} \log\Big[(2\pi e)\{(\alpha\bar{\beta} P+ \lambda_1^2 Q)- \lambda_1^2 \lambda_2^2 Q^2 +\frac{C}{D}\}\Big],\\\label{eqn34}
H(S,Z|U)   &=& \frac{1}{2} \log\Big[(2\pi e)\{\frac{\alpha\beta PQ}{\alpha\beta P+ \lambda_2^2 Q}\}\Big]+ \log\Big[(2\pi e)\{\alpha\bar{\beta} P+ N_2\}\Big]\\\label{eqn35}
H(S,Z|U, V )&=& \frac{1}{2} \log\Big[(2\pi e)\{\frac{\alpha^2 \beta \bar{\beta} P^2 Q}{\alpha^2\beta\bar{\beta} P^2+ \lambda_1^2\alpha\beta PQ+ \lambda_2^2\alpha\bar{\beta} PQ}\}\Big]+ \log[(2\pi e) N_2]
\end{IEEEeqnarray}
in which
\begin{IEEEeqnarray}{rCl}
A&=& -(\alpha \beta P+ \lambda_2^2 Q)(\alpha \bar{\beta} P+ \lambda_1 Q(K+1))^2 -(\alpha \bar{\beta} P
+ \lambda_1^2 Q)(\alpha \beta P+ \lambda_2 Q(K+1))^2 \IEEEnonumber\\
&&+ 2\lambda_1\lambda_2Q(\alpha \bar{\beta} P+ \lambda_1 Q(K+1))(\alpha \beta P+ \lambda_2 Q(K+1))\IEEEnonumber\\
B &=& P+(2K+1) Q+N_1 \IEEEnonumber\\
C &=& -(\lambda_1\lambda_2 Q)^2 (P+(2K+1)Q+N_1) -(\alpha \beta P+ \lambda_2^2 Q)(\alpha \bar{\beta} P+ \lambda_1 Q(K+1))^2 \IEEEnonumber\\
&& + 2\lambda_1\lambda_2 Q(\alpha \bar{\beta} P+ \lambda_1 Q(K+1))(\alpha \beta P+ \lambda_2 Q(K+1))\IEEEnonumber\\
D &=& (\alpha \beta P+ \lambda_2^2 Q) (P+(2K+1)Q+N_1) -(\alpha \beta P+ \lambda_2 Q(K+1))^2\IEEEnonumber\\
K&=& \sqrt{\frac{\bar{\alpha} P}{Q}},\IEEEnonumber\\
&&---------------------------------------------\IEEEnonumber
\end{IEEEeqnarray}
\end{figure*}
Taking derivatives of~$R_{e_1}$ with respect to~$\lambda_1$ and~$\lambda_2$
and setting to zero yields
\begin{IEEEeqnarray}{rCl}\label{eqn36}
\hat{\lambda}_1&=&\frac{\alpha\bar{\beta} P (K+1)}{\alpha P+N_1}\\\label{eqn37}
\hat{\lambda}_1&=&\frac{\alpha\beta P (K+1)}{\alpha P+N_1}
\end{IEEEeqnarray}
Then, we substitute these optimal variables in~\eqref{eqn29} to get
\begin{IEEEeqnarray}{rCl}\label{eqn38}
R_{e_1}(\hat{\lambda}_1, \hat{\lambda}_2)&=&\mathcal{C}\Big[\frac{\alpha P}{N_1}\Big],\\\label{eqn39}
R_{e_2}(\hat{\lambda}_1, \hat{\lambda}_2)&=&\mathcal{C}\Big[\frac{\alpha P}{N_1}\Big]-\mathcal{C}\Big[\frac{\alpha P}{N_2}\Big].
\end{IEEEeqnarray}

Next, we optimize~\eqref{eqn38}-\eqref{eqn39} with respect to~$\alpha$ and~$\beta$. In
the case~$N_1 > N_2$, the derived result of~$R_{e_2}$ is a decreasing
function with respect to~$\alpha$ and~$\bar{\beta}$. Thus, this function is maximized
at~$\alpha=\bar{\beta}= 0$, and~$ \lambda_1 = \lambda_2 = 0$. Therefore, the achievable
rates are equal to zero for this case, i.e.,~$R_{e_1} = R_{e_2} = 0$.

In the case that~$N_1 < N_2$,~$R_{e_2}$ is an increasing function
with respect to~$\alpha$ and~$\bar{\beta}$. Thus, this function is maximized
at~$\alpha^*=\bar{\beta}^*=1$. Finally, the variables~$\alpha^*=1, \beta^*=0,
\lambda_1^*=\frac{P}{P+N_1}, \lambda_2^*=0$ leads the secrecy achievable rate of Theorem 1
to
\begin{IEEEeqnarray}{rCl}\label{eqn40}
R_e^* &=& \min\{R_{e_1}(\lambda_1, \lambda_2), R_{e_2}(\lambda_1, \lambda_2)\}\IEEEnonumber\\
&=& \mathcal{C}\Big(\frac{P}{N_1}\Big)- \mathcal{C}\Big(\frac{P}{N_2}\Big).
\end{IEEEeqnarray}

We conclude that for the GPC for all cases of~$N_1$ and~$N_2$
the secrecy achievable rate is as follows
\begin{IEEEeqnarray}{rCl}\label{eqn41}
R_{e-GPC}^*=
\Big[\mathcal{C}(\frac{P}{N_1})- \mathcal{C}(\frac{P}{N_2})\Big]^+.
\end{IEEEeqnarray}

Note that the necessary condition, in Theorem 1, under which
the GPC gives the secrecy achievable rate, is satisfied and
discussed in Remark 5.

Now, based on Theorem 2 for the SPC case, we can define
\begin{IEEEeqnarray}{rCl}\label{eqn42}
V &=& X,\\\label{eqn43}
U&=& \emptyset.
\end{IEEEeqnarray}
substituting these parameters in~\eqref{eqn11} we have
\begin{IEEEeqnarray}{rCl}\label{eqn44}
&&I(V ; Y |U, S) -I(V ;Z|U, S)\IEEEnonumber\\
&=& I(X; Y |S)- I(X;Z|S)\IEEEnonumber\\
&=& H(Y |S) - H(Y |X, S) - H(Z|S) + H(Z|X, S)\IEEEnonumber\\
&=& H(X + \eta_1) - H(\eta_1) - H(X + \eta_2) + H(\eta_2)\IEEEnonumber\\
&=& \mathcal{C}(\frac{P}{N_1})- \mathcal{C}(\frac{P}{N_2}),
\end{IEEEeqnarray}
which means that the chosen parameters in~\eqref{eqn42}-\eqref{eqn43}, lead
Theorem 2 to the following secrecy achievable rate
\begin{IEEEeqnarray}{rCl}\label{eqn45}
R_{e-SPC}^*=
\Big[\mathcal{C}(\frac{P}{N_1})- \mathcal{C}(\frac{P}{N_2})\Big]^+.
\end{IEEEeqnarray}

Note that the necessary condition, in Theorem 2, under which
the SPC gives the secrecy achievable rate, is satisfied and
discussed in Remark 5.

For the converse proof, using the fact that the legitimate receiver is more capable
than the wiretapper, we can use the result of Corollary~\ref{col3}.
Thus, we should prove that choosing the Gaussian distribution for the channel input,
maximizes the achievable rate to the capacity of the wiretap cannel without channel state.
Without loss of generality, we assume that the channel is physically degraded, i.~e.,
$\eta_2=\eta_1+ \eta_2^{'}$, in which~$\eta_2^{'} \sim N(0,N_2-N_1)$. For the outer bound on the
capacity of the channel we have
\begin{IEEEeqnarray}{rCl}\label{eqn451}
\!\!\!\!\!\!\!I(X ; Y | S)&-& I(X ;Z| S)\IEEEnonumber\\
&=& I(X ; X+\eta_1)- I(X ; X+\eta_2)\IEEEnonumber\\
&=& H(X+\eta_1)-H(\eta_1)- H(X+\eta_2)+ H(\eta_2)\IEEEnonumber\\
&=& \frac{1}{2}\log\Big(\frac{N_2}{N_1}\Big)-[H(X+\eta_2)-H(X+\eta_1)].
\end{IEEEeqnarray}
Then, by substituting~$M=X+\eta_1 $ we have
\begin{IEEEeqnarray}{rCl}\label{eqn452}
\!\!\!\!\!\!\!\!\!\!\! H(X+\eta_2)&-&H(X+\eta_1)\IEEEnonumber\\
&=& H(M+\eta_2^{'})-H(M)\IEEEnonumber\\
&\stackrel{(c)}{\geq}& \frac{1}{2}\log(2^{2H(\eta_2^{'})}+ 2^{2H(M)})- H(M)\IEEEnonumber\\
&=& \frac{1}{2}\log(2\pi e (N_2 - N_1)+ 2^{2H(M)})- H(M),
\end{IEEEeqnarray}
where~$(c)$ is derived by the entropy power inequality (EPI)~\cite{bibi69}.
Moreover,~$\frac{1}{2}\log(2\pi e (N_2 - N_1)+ 2^{2H(u)})- H(u)$ is a monotonic increasing function
with respect to~$u$ and~$H(M) \leq \frac{1}{2}\log(2\pi e (P+ N_1))$. Thus, we have
\begin{IEEEeqnarray}{rCl}\label{eqn453}
H(X+\eta_2)&-&H(X+\eta_1)\IEEEnonumber\\
&\geq& \frac{1}{2}\log(2\pi e (N_2 - N_1)+ 2\pi e (P+ N_1))\IEEEnonumber\\
&&- \frac{1}{2}\log(2\pi e (P+ N_1))\IEEEnonumber\\
&=& \frac{1}{2}\log\Big(\frac{P+N_2}{P+N_1}\Big).
\end{IEEEeqnarray}
Finally, we have
\begin{IEEEeqnarray}{rCl}\label{eqn453}
I(X ; Y | S)&-& I(X ;Z| S)\IEEEnonumber\\
&\geq& \frac{1}{2}\log\Big(\frac{N_2}{N_1}\Big)- \frac{1}{2}\log\Big(\frac{P+N_2}{P+N_1}\Big)\IEEEnonumber\\
&=& \mathcal{C}(\frac{P}{N_1})- \mathcal{C}(\frac{P}{N_2}),
\end{IEEEeqnarray}
and the equality is attained by choosing~$X\sim \mathcal{N} (0, P)$.
\end{IEEEproof}

\begin{figure}
\centering
\epsfig{file=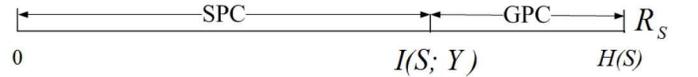,width=1\linewidth,clip=.5}
\caption{The conditions under which GPC (by substituting R.V.s as~$V = X+(\frac{P}{P+N_1})S$) and SPC (by substituting R.V.s~$V = X; U = \emptyset$)
are used in Gaussian state-dependent wiretap channel.}
\label{fig:6}
\end{figure}

\begin{remark}
We should note that GPC is used in the case
that~$ \min\{I(S;U, Y ), I(S; V, Y |U)\} \leq R_S \leq H(S)$. By
substituting the optimal parameters~$\alpha^*=1, \beta^*=0,
\lambda_1^*=\frac{P}{P+N_1}, \lambda_2^*=0$,
this condition is reduced to~$I(S; Y ) \leq
R_S \leq H(S)$. On the other hand, SPC is used when~$R_S\leq
\min\{I(S;U, Y ), I(S; V, Y |U)\}$, which by substituting R.V.s
as~\eqref{eqn42}-\eqref{eqn43}, we have~$R_S\leq I(S; Y )$.
As shown in Fig.~\ref{fig:6}, our proposed coding schemes meet the
capacity for any~$R_S$. Note that
in the Gaussian case,~$I(S; Y ) = \mathcal{C}(\frac{Q}{P+N_1})$.
\end{remark}

\begin{remark}
In GPC scheme, choosing~$\alpha^* = \bar{\beta}^*=1$ and~$\lambda_1^*, \lambda_2^*$
results in
\begin{IEEEeqnarray}{rCl}\label{eqn46}
V&=& X+\Big(\frac{P}{P+N_1}\Big)S,\IEEEnonumber\\
U&=&\emptyset,
\end{IEEEeqnarray}
and it is noticeable that the parameter~$\lambda^*_1= \frac{P}{P+N_1}$
is similar to the one chosen in dirty paper channel~\cite{bibi30} to achieve
the capacity in the state-dependent channel. This inspired the
authors to name the proposed method \textit{Secure Dirty Paper Coding} (SDPC).
\end{remark}

\section{Discussions and Conclusions}\label{Conc}
In this paper we derived two equivocation-rates for the
state-dependent wiretap channel in which the channel state
information is assumed to be known non-causally
at the transmitter. These equivocation-rates are derived from
the equivocation-rate regions reported for the cognitive interference
channel~\cite{bibi31, bibi9}. Comparing our model to
the cognitive interference channel, the channel state plays
the role of the message of the primary user.
The transmitter uses the coding schemes
previously used by the cognitive transmitter, i.e. Gel'fand-Pinsker
coding and superposition coding.
By this point of view, we derived new achievable equivocation-rates for
the state-dependent wiretap channel. Then, we showed that
the derived equivocation-rates meet the capacity of the state-dependent
wiretap channel under some conditions. As an
example, the secure capacity of a state-dependent binary
symmetric channel was considered which confirms the general
results.
Afterward, the state-dependent Gaussian wiretap channel
was studied, and our achievable equivocation-rates
lead to the capacity in Gaussian case. It was shown that the
capacity of the state-dependent Gaussian wiretap channel is
equal to the capacity of the Gaussian wiretap channel without
channel state. This result was derived using dirty paper coding
approach~\cite{bibi30}, by maximizing the equivocation-rates. The
authors called this coding scheme Secure Dirty Paper Coding.

To compare our model with the one presented in~\cite{bibi23},
we should note that in~\cite{bibi23}, the transmitter which
non-causally knows the CSI, uses this information to increase its
secrecy rate to~$R^{Chen-Vinck}=I(V; Y)-\max\{I(V;S), I(V;Z)\}$. Therein, it is assumed that the CSI is not known at
the eavesdropper. Thus, in the case that~$I(V;S)\geq I(V;Z)$, binning the codewords into~$2^{nI(V;S)}$ bins, overcomes
the channel state, and in this case the message will be kept secure by the random coding scheme. In Gaussian case, when~$I(V;S)\geq I(V;Z)$,
this approach leads to the capacity of the
point to point channel, i.~e.,~$\frac{1}{2}\log(1+\frac{P}{N_1})$. Furthermore, the output of the channel at the eavesdropper
assumed to be a degraded version of the one at the legitimate receiver, i.~e., the~$U \rightarrow (X,S) \rightarrow Y \rightarrow Z$
forms a Markov chain in~\cite{bibi23}.
But in the model considered in this paper, the CSI can be decoded at the eavesdropper.
Thus, the CSI cannot be used to improve the secrecy rate in SD-WC. It is noticeable that
due to the capability of the eavesdropper to estimate the channel,
the assumption that the CSI can be decoded at the eavesdropper sounds a little realistic.
Moreover, in our model it is not necessary to assume the channel output at the eavesdropper
to be a degraded version of the legitimate receiver's one.

\appendices
\section{Proof of the Lemma 2}\label{App:A}
The SD-WC is modeled with a CIC, i.~e., the CSI is considered as a primary
transmitter's message and thus is transmitted through
the channel; and the transmitter in SD-WC plays the role of a cognitive
transmitter who has the message of the primary one non-causally.
On the other hand, the transmitted message in SD-WC must
be confidential at the primary receiver who acts as an eavesdropper
for the cognitive transmitter's message. First, note that
in~\cite{bibi31} two confidential messages are considered in CIC
model, i.~e., the primary and the cognitive receivers act as
eavesdroppers for each other's message. Here, we just consider
the cognitive transmitter's message to be confidential at the
primary receiver. Hence, we reduce the rate region of~\cite{bibi31} to the CIC with
one confidential message by excluding extra secrecy condition.
Thus, we have the following lemma.

\begin{figure}
\centering
\epsfig{file=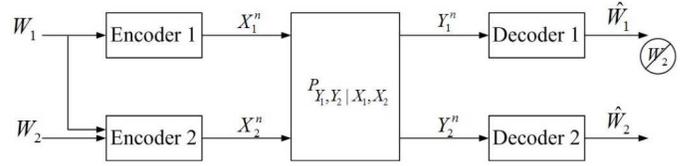,width=1\linewidth,clip=.5}
\caption{Cognitive interference channel with a confidential messages.}
\label{fig:7}
\end{figure}

\begin{lemma}{~\cite[Theorem 1]{bibi31}}
The set of the rates~$(R_1,R_{2_a},R_c,R_{e_2} )$ satisfying
\begin{IEEEeqnarray}{rCl}\label{eqn48}
R_1     &\leq& I(X_1; U, Y_1 | T), \\
R_{2a} &\leq& I(V ; Y_2 | U, T) - I(V ;X_1 | U, T), \\
R_2     &\leq& I(U, V ; Y_2 | T) - I(U, V ;X_1 | T), \\
R_1 + R_c &\leq& I(X_1,U; Y_1 | T), \\
R_{e_2} &\leq& I(V ; Y_2 | U, T) - I(V ;X_1, Y_1 | U, T),
\end{IEEEeqnarray}
is achievable for CIC with a confidential message.

Now, comparing the SD-WC (Fig.~\ref{fig:1}) with the CIC (Fig.~\ref{fig:7}),
we can derive a new achievable rate for the SD-WC. We should
note that the message of the primary transmitter plays the role of the channel sate in
SD-WC. Since the eavesdropper is not forced to decode the
channel state in SD-WC, we should relax the terms contain~$R_1$ which
is related to the rate of the primary transmitter's message (the channel state in SD-WC).
Thus, by setting
\begin{IEEEeqnarray}{rCl}
W_2 = W,\IEEEnonumber
R_1 = R_S,R_{2a} = R_1,\IEEEnonumber\\
R_c = R_2,R_{e_2} = R_e,\IEEEnonumber\\
R = R_1 + R_2,\IEEEnonumber\\
Y_1 = Z, Y_2 = Y,\IEEEnonumber\\
X_1 = S,X_2 = X,
\end{IEEEeqnarray}
and relaxing the rates (48) and (51), we derive the equivocation-rate
of Lemma~\ref{lem2}. Finally, we remark that we can derive the equivocation-rate
of Lemma~\ref{lem2} directly by introducing the codebook generation,
encoding and decoding schemes, error analysis and equivocation
computation similar to the one presented in~\cite{bibi31}.
\end{lemma}

\section{Proof of the Lemma 3}\label{App:B}
Consider the CIC with one confidential message (Fig.~\ref{fig:7}).
Using the SPC, the achievable equivocation-rate region for this
channel is derived by~\cite{bibi9} as follows.
\begin{lemma}~\cite[Theorem 1]{bibi9}
The set of the rates~$(R_1,R_2,R_e)$ satisfying
\begin{IEEEeqnarray}{rCl}\label{eqn53}
R_1 &\leq& \min\{I(U,X_1; Y ), I(U,X_1;Z)\}\\\label{eqn54}
R_2 &\leq& I(U, V ;Z | X_1), \\\label{eqn55}
R_1 + R_2 &\leq& \min\{I(U,X_1; Y ), I(U,X1;Z)\}\IEEEnonumber\\
&&  + I(V ;Z | U,X_1), \\\label{eqn56}
R_e &\leq& I(V ;Z | U,X_1) - I(V ; Y | U,X_1),
\end{IEEEeqnarray}
is achievable for CIC with a secret message.

Now, by substituting~$R_2 = R,X_1 = S$ and relaxing
the rates~\eqref{eqn53} and~\eqref{eqn55} which correspond to the primary
user (channel state in our setting) the achievability of the
equivocation-rate of Lemma~\ref{lem3} is proved.
\end{lemma}

\section{The Converse Proof of Theorem 3}\label{App:C}
The converse proof of Theorem~\ref{thm3} is derived as
follows. Consider the rate-pair~$(R,R_e)$ to be achievable. Then,
we have
\begin{IEEEeqnarray}{rCl}\label{eqn57}
NR &\stackrel{(a)}{\leq}& I(W; Y^N) - I(W;Z^N, S^N) + \epsilon\IEEEnonumber\\
   &{\leq}& I(W; Y^N| S^N) - I(W;Z^N|S^N) + \epsilon\IEEEnonumber\\
   &\leq& \sum_{i=1}^{N}  I(W; Y_i|Y^{i-1}, S^N) - I(W;Z^i|Z_{i+1}^N, S^N) + \epsilon\IEEEnonumber\\
   &\stackrel{(b)}{\leq}& \sum_{i=1}^{N} I(W, Z_{i+1}^N; Y_i|Y^{i-1}, S^N) \IEEEnonumber\\
   &&- I(W, Y^{i-1};Z^i|Z_{i+1}^N, S^N) + \epsilon\IEEEnonumber\\
   &\stackrel{(c)}{\leq}& \sum_{i=1}^{N} I(W; Y_i|Y^{i-1}, Z_{i+1}^N, S^N) \IEEEnonumber\\
   &&- I(W;Z^i| Y^{i-1}, Z_{i+1}^N, S^N) + \epsilon\IEEEnonumber\\
   &\stackrel{(d)}{=}& \sum_{i=1}^{N} I(V_i; Y_i|U_i, S_i) - I(V_i;Z_i|U_i, S_i) + \epsilon\IEEEnonumber\\
   &\stackrel{(e)}{=}&I(V; Y|U, S) - I(V;Z|U, S) + \epsilon\IEEEnonumber\\
\end{IEEEeqnarray}
in which
~$(a)$ follows from the Fano's inequality and the
fact that~$I(W;Z^N, S^N$) tends to zero for~$N\rightarrow \infty$;
~$(b)$ and~$(c)$ follow from the Csisz\'{a}r sum identity~\cite{bibi69};
~$ (d)$ is derived by substituting
the random variables~$ U_i = (Y^{i-1},Z^N_{i+1}, S^{i-1}, S^N_{i+1}), V_i =
(W,U_i)$, and~$ (e)$ follows by defining a time-sharing random
variable~$ Q$ and defining~$ U = (U_Q,Q), V = (V_Q,Q), Y = YQ,$
and~$ Z = Z_Q$. This completes the proof.

\bibliography{thesisbib}

\begin{thebibliography}{10}

\bibitem{bibi23}
Y.~Chen and H.~Vinck,
\newblock ``Wiretap channel with side information,''
\newblock {\em IEEE Trans. Inf. Theory}, vol. 54, no. 1, pp. 395--402, Jan.
  2008.

\bibitem{bibi21}
C.~Mitrpant, H.~Vinck, and Y.~Luo,
\newblock ``An achievable region for the \textsc{G}aussian wiretap channel with
  side information,''
\newblock {\em IEEE Trans. Inf. Theory}, vol. 52, no. 5, pp. 2181--2190, May
  2006.

\bibitem{bibi69}
A.~El~Gamal and Y.-H. Kim,
\newblock {\em Network Information Theory},
\newblock Cambridge University Press, 2011.

\bibitem{bibi79}
Y.~Liang, H.~V. Poor, and S.~Shamai,
\newblock {\em Information Theoretic Security},
\newblock Now Pub. Inc., 2009.

\bibitem{bibi76}
H.~G. Bafghi, M.~Mirmohseni, B.~Seyfe, and M.~R. Aref,
\newblock ``On the secrecy of the cognitive interference channel with partial
  channel states,''
\newblock {\em Submitted to Trans. on Emerging Telecom. Tech.: available on
  arXiv:1511.07168v1 [cs.IT] 23 Nov 2015}, 2016.

\bibitem{bibi7}
A.~Wyner,
\newblock ``The wire-tap channel,''
\newblock {\em Bell Syst. Tech. J.}, vol. 54, no. 8, pp. 1355--1387, Jan. 1975.

\bibitem{bibi28}
C.~E. Shannon,
\newblock ``Communication theory of secrecy systems,''
\newblock {\em Bell Syst. Tech. J.}, vol. 28, pp. 656--715, Oct. 1949.

\bibitem{bibi39}
C.~E. Shannon,
\newblock ``Channels with side information at the transmitter,''
\newblock {\em J. Res. Devel.}, vol. 2, pp. 289--293, 1958.

\bibitem{bibi22}
S.~I. Gel'fand and M.~S. Pinsker,
\newblock ``Coding for channel with random parameters,''
\newblock {\em Probl. Inf. Theory}, vol. 9, no. 1, pp. 19--31, 1980.

\bibitem{bibi30}
M.~H.~M. Costa,
\newblock ``Writing on dirty paper,''
\newblock {\em IEEE Trans. Inf. Theory}, vol. IT-29, no. 3, pp. 439–--441, May
  1983.

\bibitem{bibi75}
Y.~Kim, A.~Sutivong, and T.~Cover,
\newblock ``State amplification,''
\newblock {\em IEEE Trans. Inf. Theory}, vol. 54, no. 5, pp. 1850--1859, May
  2008.

\bibitem{bibi77}
N.~Merhav and S.~Shamai,
\newblock ``Information rates subjected to state masking,''
\newblock {\em IEEE Trans. Inf. Theory}, vol. 53, no. 6, pp. 2254--2261, June
  2007.

\bibitem{bibi40}
A.~Khisti, S.~Diggavi, and G.~Wornell,
\newblock ``Secret-key agreement with channel state information at the
  transmitter,''
\newblock {\em IEEE Trans. Forens. and Sec.}, vol. 6, no. 3, pp. 672--681,
  Sept. 2011.

\bibitem{bibi42}
Y.~Chia and A.~El-Gamal,
\newblock ``Wiretap channel with causal state information,''
\newblock {\em IEEE Trans. Inf. Theory}, vol. 58, no. 5.

\bibitem{bibi31}
H.~G. Bafghi, S.~Salimi, B.~Seyfe, and M.~R. Aref,
\newblock ``Cognitive interference channel with two confidential messages,''
\newblock in {\em Int. Symp. on Inf. Theory and Applic. (ISITA)}, Taichung,
  Taiwan, 2010, pp. 952--956.

\bibitem{bibi9}
Y.~Liang, A.~Somekh-Baruch, H.~V. Poor, S.~Shamai, and S.~Verd\'{u},
\newblock ``Capacity of cognitive interference channels with and without
  secrecy,''
\newblock {\em IEEE Trans. Inf. Theory}, vol. 55, no. 2, pp. 604--618, Feb.
  2009.

\bibitem{bibi13}
I.~Maric, A.~Goldsmith, G.~Kramer, and S.~Shamai,
\newblock ``On the capacity of interference channels with one cooperating
  transmitter,''
\newblock {\em European Trans. on Telecomm.}, vol. 19, no. 4, pp. 405–--420,
  2008.

\bibitem{bibi34}
T.~M. Cover and J.~A. Thomas,
\newblock {\em Elements of Information Theory},
\newblock John Wiley and Sons, Inc., 2nd edition, 2006.

\end{thebibliography}
\bibliographystyle{IEEE}

\end{document}